# Scalar – Tensor gravity with scalar – matter direct coupling and its cosmological probe

Jik Su Kim

*Pyongyang Astronomical Observatory, Academy of Sciences, Pyongyang, DPR Korea*

Chol Jun Kim, Sin Chol Hwang, Yong Hae Ko,

*Department of Physics, **Kim Il Sung** University, Pyongyang, DPR Korea*

Abstract

SNIa and CMB datasets have shown both of evolving Newton's "constant" and a signature of the coupling of scalar field to matter. These observations motivate the consideration of the scalar-matter coupling in Jordan frame in the framework of scalar-tensor gravity. So far, majority of the works on the coupling of scalar to matter has been performed in Einstein frame in the framework of minimally coupled scalar fields. In this paper, we generalize the original scalar-tensor theories of gravity by introducing a direct coupling of scalar to matter in the Jordan frame. The combined consideration of both evolving Newton's constant and scalar-matter coupling using the recent observation datasets, shows features different from the previous works. The analysis shows a vivid signature of the scalar-matter coupling. The variation rate of the Newton's constant is obtained rather greater than that determined in the previous works.

## I. INTRODUCTION

Cosmological observation datasets are opening a wide possibility of test of the various cosmological models. Nesseris and Perivolaropoulos[1] have shown that Gold dataset of SNIa yielded some evidence of the scalar–tensor property of gravitation. Making use of Gold dataset of SNIa[2], they have found the Newton's gravitation constant to be evolved.

On the other hand, Majerotto, Sapone and Amendola[3] and Guo, Ohta and Tsujikawa [4] have found that combined analysis of SNLS, CMB, and BAO datasets showed a signature of direct scalar–matter coupling. The latter, however, had been based on the background of Einstein tensor gravity.

The above both analyses are making use of almost the same observation datasets, but their results are quite contradictory, so we cannot be sure which of these models should be accepted.

As is well known, when the coupling of scalar to background space-time vanishes the gravity returns to Einstein tensor gravity. Therefore, if we want to elucidate whether both of scalar–background space–time and scalar–matter couplings do exist or not, one should construct a more inclusive model than the previous ones[1, 3, 4].

We construct a model involving a scalar–matter direct coupling in the scalar–tensor gravity. Generalized theory of scalar-tensor gravity has already been studied in [5, 6, 7].

We investigate here the theory in Jordan frame as a physical frame rather than Einstein frame. To test our model we use a new heterogeneous compilation "Union" of Type Ia supernovae[8], and the distance parameter of baryon acoustic oscillation[9]. Combining the above cosmological observation datasets we obtain a definite constraint on the cosmological parameters, especially, a variation rate of gravitation "constant" and a parameter characterizing the direct scalar–matter coupling.

In Sect.II we construct a general formalism for the scalar–tensor gravity with direct coupling of scalar field to matter. And we apply the dynamical equations to the expanding FRW model of the Universe. In Sect. III, making use of cosmological observation datasets of SNIa, and BAO, we find observational constraints on the coupling parameters. Finally, in Sect. IV, we summarize results and discuss several implications of the results.

## II. GENERAL FORMALISM FOR COUPLED SCALAR-TENSOR TEHORY OF GRAVITY

### A. General equations

The general formalism of the metric scalar-tensor theories of gravity is described in Refs. [5,6,7,10,11,12,13]. We generalize it including the scalar-matter coupling in Lagrangian, thus making the theory non-metric. We consider the action in Jordan frame

$$S = \int d^4 x \sqrt{-g} \left[ \frac{1}{2} \left( F(\varphi) R - g^{\mu\nu} \partial_\mu \varphi \partial_\nu \varphi \right) - V(\varphi) + C(\varphi) L_m^{(0)}(\Psi; g_{\mu\nu}) \right] \qquad (1)$$

where the action is characterized by three functions: scalar-curvature coupling function $F(\phi)$, scalar-matter coupling function $C(\phi)$, and self-interaction potential $V(\phi)$. $L_m^{(0)}$ describes the matter sector of the Lagrangian without scalar coupling, $\Psi$ expresses generically the matter fields and $F(\phi) > 0$ to ensure positivity of the energy of graviton. Superscript 0 in the matter Lagrangian denotes that it does not include scalar field $\phi$, but observations measure the coupled energy corresponding to $C(\phi) L_m^{(0)}(\Psi; g_{\mu\nu}) = L_m(\Psi, \phi; g_{\mu\nu})$.

The kinetic term of the scalar field in action(1) may have a factor $\omega(\phi)$, but in the case of satisfying the observational constraint on the current coupling strength $\alpha_0$ it can always be set equal to unity by a redefinition of the field $\phi$ [14]. In the action (1) we set $8\pi G_N = 1$ where $G_N$ is Newton constant, so $F(\phi)$, $C(\phi)$ and scalar field $\phi$ are dimensionless, and in this unit energy density and potential $V$ have dimension (length)$^{-2}$. The coupling function $C(\phi)$ makes the theory non-metric, and thereby presupposes the violation of WEP and the change with time of the fundamental constant, such as, e.g. the fine-structure constant $\alpha_{EM}$ of magnitude $\Delta\alpha_{EM}/\alpha_{EM} = (-0.72 \pm 0.18)\times 10^{-5}$ for the red shift interval $z \approx 0.5 - 3.5$ [11], and such observations, besides the motivation mentioned in Sect. I, justify the consideration of the direct scalar-matter coupling.

Many authors treat the scalar-matter coupling including it in Lagrangian $L_m(\Psi, \phi; g_{\mu\nu})$ but we represent it explicitly separating the factor $C(\phi)$ to emphasize its role. Casas et al. have explicitly separated the coupling [12].

The variations of action (1) with respect to metric tensor $g^{\mu\nu}$ and scalar field $\phi$ yield dynamical equations for gravity and scalar field

$$\left(R_{\mu\nu} - \frac{1}{2}g_{\mu\nu}R\right)F(\phi) = C(\phi)T^{(0)}_{\mu\nu} + T_{\mu\nu(\phi)}, \tag{2}$$

$$2\Box\phi = -\frac{dF(\phi)}{d\phi}R + 2\frac{dV(\phi)}{d\phi} - 2\frac{dC(\phi)}{d\phi}T^{(0)}, \tag{3}$$

$$T_{\mu\nu(\phi)} = \phi_{,\mu}\phi_{,\nu} - \frac{1}{2}g_{\mu\nu}\phi_{,\alpha}\phi^{,\alpha} + F(\phi)_{,\nu;\mu} - g_{\mu\nu}\Box F(\phi) - g_{\mu\nu}V(\phi), \tag{4}$$

where $\Box = \frac{1}{\sqrt{-g}}\partial_\mu(\sqrt{-g}\partial^\mu)$, matter energy-momentum tensor is defined by $T^{(0)}_{\mu\nu} = -\frac{2}{\sqrt{-g}}\frac{\partial\sqrt{-g}L^{(0)}_m}{\partial g^{\mu\nu}}$, and $T^{(0)} = T^{(0)\mu}{}_\mu$. The equation of scalar field (3) can be expressed in other form using the trace of Eq. (2) and it reads

$$2\Theta(\phi)\Box\phi = \left[C(\phi)\frac{dF(\phi)}{d\phi} - 2\frac{dC(\phi)}{d\phi}F(\phi)\right]T^{(0)} - \frac{d\Theta(\phi)}{d\phi}\phi_{,\alpha}\phi^{,\alpha} - 4V(\phi)\frac{dF(\phi)}{d\phi} + 2\frac{dV(\phi)}{d\phi}F(\phi) \tag{5}$$

where $2\Theta(\phi) \equiv 2F(\phi) + 3\left(\frac{dF(\phi)}{d\phi}\right)^2$. Taking covariant derivative of Eq. (2) and using Eq. (3) we find energy-momentum conservation equation for the matter

$$T^{(0)\nu}{}_{\mu;\nu} = \frac{d\ln C(\phi)}{d\phi}\left(\delta^\nu_\mu T^{(0)} - T^{(0)\nu}{}_\mu\right)\phi_{,\nu}, \tag{6}$$

This form of the conservation equation is different from the conventional expression on the conservation law for the matter. Expressing the energy-momentum tensor by

$$T_{\mu\nu} = C(\phi)T^{(0)}_{\mu\nu}, \tag{7}$$

we can restore the conventional conservation equation

$$T^\nu_{\mu;\nu} = \frac{d\ln C(\phi)}{d\phi}T\phi_{,\mu} = \frac{dC(\phi)}{d\phi}T^{(0)}\phi_{,\mu}, \tag{8}$$

where $T_{\mu\nu} = -\frac{2}{\sqrt{-g}}\frac{\partial\sqrt{-g}L_m}{\partial g^{\mu\nu}}$ and $T \equiv T^\mu{}_\mu$. Eq. (8) resembled the equation $T^\mu_{\nu(m);\mu} = C_A T_{(m)} \phi_{,\nu}$ in [13] where $C_A$ corresponds to our $d\ln C/d\phi$. However, [13] considered this coupling in the Einstein frame and $C_A$ is constant, whereas, in our case, $d\ln C/d\phi$ is not constant and an arbitrary function of $\phi$, so the coupling function $C(\phi)$ represents most general coupling. Casas et al.[12] have investigated the direct scalar-matter coupling in the Jordan frame in the framework of Brans-Dicke theory of gravity, and their coupling function is $C(\phi) = \phi^\sigma$, so $\dfrac{d\ln C(\phi)}{d\phi}\phi_{,\mu} = \sigma\dfrac{\phi_{,\mu}}{\phi}$ .

### B. The coupled scalar-tensor theory in expanding Universe

In this subsection, we apply the general equations to the expanding Universe and consider several problems of the theory in the presence of the direct scalar-matter coupling.

We consider a flat Friedmann-Robertson-Walker Universe whose metric in the Jordan frame is given by

$$ds^2 = -dt^2 + a^2 dx^2 \tag{9}$$

where $t$ is cosmic time. To compare the outcome of all the computations with observations we perform the computation in the Jordan frame. In the following, matter will be described as a perfect fluid, so its energy-momentum tensor takes the form as

$$T_{\mu\nu} = (\rho + P)u_\mu u_\nu + g_{\mu\nu} P \tag{10}$$

where $u_\mu = dx_\mu/ds$ is the components of four velocity of the matter in Jordan frame. Taking $G_{00} = 3(\dot{a}/a)^2$ and $G_{ii} = -2(\ddot{a}/a) - (\dot{a}/a)^2$ the dynamical equations derived from Eqs (2) and (3) are given by

$$3F(\phi)H^2 = C(\phi)\rho^{(0)} + \frac{1}{2}\dot{\phi}^2 + V(\phi) - 3H\dot{F}(\phi) , \tag{11}$$

$$F(\phi)\dot{H} = -\frac{1}{2}C(\phi)\left(\rho^{(0)} + P^{(0)}\right) - \frac{1}{2}\dot{\phi}^2 - \frac{1}{2}\ddot{F}(\phi) + \frac{1}{2}H\dot{F}(\phi) , \tag{12}$$

$$\ddot{\phi} + 3H\dot{\phi} = \frac{3dF(\phi)}{d\phi}\left(\dot{H} + 2H^2\right) - \frac{dV(\phi)}{d\phi} + \frac{dC(\phi)}{d\phi}T^{(0)} , \tag{13}$$

where subscript 0 stands for the quantity corresponding to Lagrangian $L_m^{(0)}$ in the action (1). From Eq. (6) we get immediately

$$\dot{\rho}^{(0)} + 3H\left(\rho^{(0)} + P^{(0)}\right) = 0 . \tag{14}$$

In spite of the direct coupling of the scalar to the matter, the matter density $\rho^{(0)}$ which does not take into account the coupling is conserved. However, as mentioned above, the observations measure the coupled matter energy corresponding to $C(\phi)L_m^{(0)} = L_m(\phi)$, one can define the coupled matter density $\rho_m(\phi) = C(\phi)\rho^{(0)}$, and corresponding conservation equation (8) yields

$$\dot{\rho}_m(\phi) + 3H(\rho_m(\phi) + P_m(\phi)) = \frac{d\ln C(\phi)}{d\phi}\dot{\phi}\rho_m(\phi) \quad . \tag{15}$$

This equation shows that the matter conservation is unaffected by scalar-curvature coupling function $F(\phi)$ and the direct coupling of scalar to the matter violates the matter conservation law. In the previous works[3, 4, 9, 12, 13, 15, 16, 17], the matter conservation equation in the presence of the scalar-matter coupling is customary to be written in the form

$$\dot{\rho}_m + 3H(\rho_m + P_m) = \Gamma\rho_m = \delta H\rho_m \ . \tag{16}$$

Comparing with Eq. (15) one can find

$$\Gamma = \frac{\dot{C}(\phi)}{C(\phi)} = \frac{d\ln C}{d\phi}\dot{\phi} \ . \tag{17}$$

From the relation $\Gamma = \delta H$, (16), we get immediately the following meaningful relation

$$\frac{\dot{C}}{C} = \delta\frac{\dot{a}}{a} \rightarrow C \sim a^\delta \ . \tag{18}$$

This simple relation implies that in positive $\delta$ the coupling function $C(\phi)$ increased with expansion of the Universe and thereby particle mass increases, while in negative $\delta$ it reduces with time.

Let us define the energy density and pressure of the scalar field. It is straightforward to deduce them from the definition of energy-momentum tensor for scalar field (4).

$$\rho_\phi = T_{00(\phi)} = \frac{1}{2}\dot{\phi}^2 + V(\phi) - 3H\dot{F}(\phi) \ , \tag{19}$$

$$P_\phi = \frac{1}{3}T_{ij(\phi)}\delta^{ij} = \frac{1}{2}\dot{\phi}^2 - V(\phi) + \ddot{F}(\phi) + H\dot{F}(\phi) \ . \tag{20}$$

Multiplying the Eq. (13) by $\dot{\phi}$ and taking into account of the definition (19) and (20) we get immediately the equation of energy conservation for the scalar field

$$\dot{\rho}_\phi + 3H(\rho_\phi + P_\phi) = -\frac{d\ln C(\phi)}{d\phi}\dot{\phi}\rho_m(\phi) \ . \tag{21}$$

The conservation equations for the matter and the scalar field (15) and (21) are derived in the Jordan frame, but, as mentioned above, the conservation equations in the Jordan frame have the same form as in the Einstein frame, and actually, almost all the previous studies of the scalar-matter coupling in Einstein frame are giving the same equations as (15) and (21). The difference is merely that the parameter $\Gamma$ is expressed by $d(\ln C)/dt$, but this is an elucidation of the implication of the parameter $\Gamma$. Both of the conservation equations (15) and (21) are giving the conservation of the combined system <matter + scalar field>.

Bellow in Sect. III the coupling parameter $\Gamma$ or $\delta$ will be determined in the Jordan frame making use of the recent SNIa and BAO datasets. Previous studies concerning the direct scalar-matter coupling have been performed in the Einstein frame[3, 4]. Recent observations show a signature of the scalar-tensor gravity[1]. Therefore, to reexamine the scalar-matter coupling in the Jordan frame deserves further attention.

## III. OBSERVATIONAL CONSTRAINTS ON THE COUPLING FUNCTIONS $F(z)$ AND $C(z)$

Recent observations of SNIa standard candles[2, 8, 18, 19], CMB anisotropy and the baryon acoustic oscillations (BAO) in the Sloan Digital Sky Survey (SDSS) luminous galaxy sample [20] provide a new prospect to determine the cosmological parameters.

Nesseris and Perivolaropouls[1], using the Gold dataset of SNIa, have shown an observational evidence for the evolving Newton's "constant" and thereby the scalar-tensor character of the gravitation. They have used the Gold dataset to determine the parameters of simple polynomial expressions for the functions $H(z)$ and $G(z)$ of the form

$$H^2(z) = H_0^2 \left\{ \Omega_{0m}(1+z)^3 + a_1(1+z) + a_2(1+z)^2 + (1 - \Omega_{0m} - a_1 - a_2) \right\} \quad , \tag{22}$$

$$G(z) = G_0 (1 + az^2). \tag{23}$$

In the determination of evolving $G(z)$, they have made use of the fact that the peak luminosity of SNIa is proportional to the mass of nickel synthesized which is a fixed fraction of the Chadrasekhar mass $M_{ch}$ varying as $M_{ch} \sim G^{-\frac{3}{2}}$. Therefore, the SNIa peak luminosity would vary like $L \sim G^{-\frac{3}{2}}$ and the corresponding SNIa absolutely magnitude evolves like

$$M - M_0 = \frac{15}{4} \log \frac{G}{G_0} \tag{24}$$

where the subscript 0 denotes the local values of $M$ and $G$. On the other hand, the Newton's constant enters the Friedmann equation. In the scalar-tensor theories, it relates to the scalar-curvature coupling function $F$ by a relation[3, 9]

$$G(z) = G_0 \frac{1}{F} \frac{2F + 4(dF/d\phi)^2}{2F + 3(dF/d\phi)^2} \approx \frac{G_0}{F} \quad . \tag{25}$$

The luminosity distance, therefore, involves the evolving gravitation constant

$$d_L(z) = (1+z) \int_0^z dz' \left( \frac{G_0}{G(z')} \right)^{\frac{1}{2}} \frac{1}{H(z')} \quad . \tag{26}$$

Then, the theoretical magnitude of observed SNIa in the context of scalar-tensor gravity is given by

$$m^{th}(z) = M_0 + 5 \log d_L(z) + \frac{15}{4} \log \frac{G(z)}{G_0}. \tag{27}$$

Substituting the polynomial expression (22) and (23) into (26) and (27) and fitting $G(z)$ and $H(z)$ to the Gold dataset of SNIa by $\chi^2$-minimization, Nesseris and Perivolaropoulos[1] have determined the parameters $a_1$, $a_2$, and $a$ in Eqs (22) and (23).

In this paper, we follow the same route as in [1], but several modifications must be appended in order to take into account of the scalar-matter coupling. In addition, at variance with (23), we take a new expression of Newton's constant as follows

$$G(z) = G_0(1 + bz + az^2) \tag{28}$$

which includes a linear term. The inclusion of the linear term in the expression (28) is justified by the fact that in the context of scalar-tensor theories of gravity the Newton's constant has the form[3, 9]

$$G(z) = \frac{G_0}{F}(1 + \alpha^2) \tag{29}$$

$$\alpha^2 = \frac{(dF(\phi)/d\phi)^2}{2F(\phi) + 3(dF(\phi)/d\phi)^2} . \tag{30}$$

The result of the Doppler tracking of the Cassini spacecraft provides an observational constraint on the current value of the coupling constant $\alpha_0^2$ [19]

$$\alpha_0^2 \leq -(1.05 \pm 1.15) \times 10^{-5} . \tag{31}$$

This implies that we can not yet entirely disregard the existence of the scalar partner in the gravitational interaction between two bodies, and that as we can see in Eq. (30), non-vanishing $\alpha^2$ implies the non-vanishing $dF(\phi)/d\phi|_{z=0} \neq 0$. Actually, from the relation (28)

$$F(z) = \frac{G_0}{G(z)} = (1 + bz + az^2)^{-1} \tag{32}$$

we have $F'(z = 0) = b$, whereas the expression (23) yields $F'(z = 0) = 0$. The expression (28) or (32) is immediately related to the rate of time varying of the Newton's constant as follows

$$\dot{G}/G|_{z=0} = -bH_0 . \tag{33}$$

In point of fact, recent observations[18] show

$$|\dot{G}/G| \leq 9 \times 10^{-13} \, yr^{-1} \tag{34}$$

which we can compare with (33). Thus, only the linear term in Eq. (28) allows to estimate the time varying gravitational constant in the present epoch.

To include the coupling function $C(\phi)$ in the Hubble parameter $H$, we take an expression of $H^2(z)$ different from (22). From Eqs (11) and (19) we have the Friedmann equation as follows

$$H^2 = \frac{1}{3F(\phi)}(C(\phi)\rho^{(0)} + \rho_\phi) . \tag{35}$$

On the other hand, from Eq. (16) we have

$$\rho_m = \rho_{m0} a^{-3+\delta} = \rho_{m0}(1+z)^{3-\delta}. \tag{36}$$

Substituting Eq. (27) into Eq. (21) we get the following solution for the energy density of the scalar field

$$\rho_\phi = \rho_{\phi 0}(1+z)^{3(1+w_\phi)} + \rho_{m0}\frac{\delta}{\delta + 3w_\phi}\left[(1+z)^{3(1+w_\phi)} - (1+z)^{3-\delta}\right]. \tag{37}$$

Putting Eqs (36) and (37) in Eq. (35) and remembering $\rho_m = C(\phi)\rho^{(0)}$ above Eq. (15), we obtain final expression on the Hubble parameter

$$\frac{H^2(z)}{H_0^2} = \frac{1}{F(z)} E^2(z), \tag{38}$$

$$E^2(z) \equiv (1-\Omega_{m0})(1+z)^{3(1+w_\phi)} + \frac{\Omega_{m0}}{\delta + 3w_\phi}\left[\delta(1+z)^{3(1+w_\phi)} + 3w_\phi(1+z)^{3-\delta}\right] \tag{39}$$

where $\Omega_{m0} = \rho_{m0}/3F_0 H_0^2$. Thus, in the scalar-tensor gravity, the Hubble parameter $H^2(z)$ contains the coupling function $F(z)$. Therefore, to separate the function $F(z)$, we expressed the expression (39) as $E^2(z)$. We have in Eqs (38) and (39) three free parameters $(\delta, w_\phi, \Omega_{m0})$ when we compare the theory with observations. Our model thus has five free parameters $(a, b, \delta, w_\phi, \Omega_{m0})$ in all. In order to reduce the number of the parameters to be determined we set the density parameter of matter $\Omega_{m0}$ to equal 0.27 [21, 22].

As a SNIa dataset, we use the "Union" compilation of 307 SNIa [8]. The Original "Union" compilation consists of 414 SNIa and it reduces to 307 SNIa after selection cuts. This "Union" compilation includes the recent large samples of SNIa from SNLS and ESSENCES Survey, the older datasets, as well as the recently extended dataset of distant supernovae observed with HST.

To unify the various heterogeneous compilations, a single consistent and blind analysis procedure is used for all the various SNIa subsamples, and a new procedure is implemented that consistently weights the heterogeneous datasets and reject outliers.

The theoretical distance modulus is defined from (27)

$$\mu^{th}(z_i) = m^{th}(z_i) - M_0 = 5\log d_L(z_i) - \frac{15}{4}\log F(z_i), \tag{40}$$

where the luminosity distance $d_L(z_i)$ is expressed, instead of (26), by

$$d_L(z_i) = (1+z_i)H_0^{-1}\int_0^{z_i}\frac{F(z')^{1/2}}{E(z')}dz'. \tag{41}$$

Thus, the theoretical model parameters are determined by minimizing the quantity

$$\chi_{SN}^2(a,b,\delta,\omega_\phi) = \sum_{i=1}^{N}\frac{[\mu^{obs}(z_i) - \mu^{th}(z_i; a,b,\delta,\omega_\phi)]^2}{\sigma_{tot}^2 + \sigma_{sys}^2 + \sum_{ij}c_i c_j C_{ij}}, \tag{42}$$

where $\sigma_{tot}$ represents an astrophysical dispersion obtained by adding in quadrature the dispersion due to lensing, $\sigma_{lens}$, the uncertainty in the Milky Way dust extinction correction and a term reflecting the uncertainty due to host galaxy peculiar velocities 300km/s. The dispersion term $\sigma_{sys}$ contains an observed sample - dependent dispersion due to possible unaccounted for systematic errors. The sum in the denominator represents the statistical uncertainty as obtained from the light-curve fit with $C_{ij}$ representing the covariance matrix of fit parameters: peak magnitudes, color and stretch and $c_i = (1, \alpha, -\beta)$ are the corresponding correction parameters.

When we use only the SNIa dataset in the model fitting, we find that Union data give a weak constraint on the coupling parameter $\delta$ as in [19], so we perform the combined analysis adding BAO datasets.

In this paper, we do not use CMB anisotropy data because the CMB anisotropy includes the information of epoch separated from now by $z = 1089$, but the datasets of SNIa and BAO only range $0 < z \leq 1$, so our parameterization (28) for varying gravitational constant does not suit for CMB. In the following paper we shall introduce a new parameterization for CMB data.

To avoid degeneracies intrinsic to the distance fitting methods we can consider also the effect of the baryon acoustic peak of the large scale correlation function at 100h$^{-1}$Mpc separation detected by the SDSS team[20]. The position of the acoustic peak is related to the distance parameter.

$$A = \frac{\sqrt{\Omega_{m0}}}{z_{BAO}} \left[ \frac{z_{BAO} F(z_{BAO})^{1/2}}{E(z_{BAO})} \left( \int_0^{z_{BAO}} \frac{F(z')^{1/2}}{E(z')} dz' \right)^2 \right]^{1/3}, \qquad (43)$$

which takes value $A_0 = 0.469 \pm 0.017$ and $z_{BAO} = 0.35$ [3]. The baryon acoustic peak is taken into account by adding

$$\chi^2_{BAO}(a, b, \delta, w_\phi, \Omega_{m0}) = \frac{(A_0 - A)^2}{\sigma_A^2} \qquad (44)$$

to the above $\chi^2_{SN}$ and $\chi^2_{CMB}$, where $\sigma_A$ is the error of $A_0$.

We perform a best fit analysis with minimization of total $\chi^2$:
$$\chi^2 = \chi^2_{SN} + \chi^2_{BAO}. \qquad (45)$$

The $\chi^2$-function (45) has four independent parameters $(a, b, \delta, w_\phi)$. Through the minimization of the $\chi^2$-function (45) we obtained best fit values of the parameters (Table1).

| parameters | Union($N_{SN} = 307$) + BAO |
|---|---|
| $a$ | $-0.13^{+0.16}_{-0.14}$ |
| $b$ | $-0.04^{+0.19}_{-0.23}$ |
| $\delta$ | $+1.01^{+10.09}_{-2.32}$ |
| $w_\phi$ | $-0.96^{+0.28}_{-0.23}$ |

Table 1. Best fit values of parameters in Jordan frame.

According to formula(33), the best fit value of parameter *b* makes it possible to determine the rate of time-variation of the gravitation constant *G*. It gives

$$\dot{G}/G|_{z=0} = 2.86 \times 10^{-12} yr^{-1} ,\qquad(46)$$

where we have taken $H_0 = 70 km/s \cdot Mpc = 7.15 \times 10^{-11} yr^{-1}$.

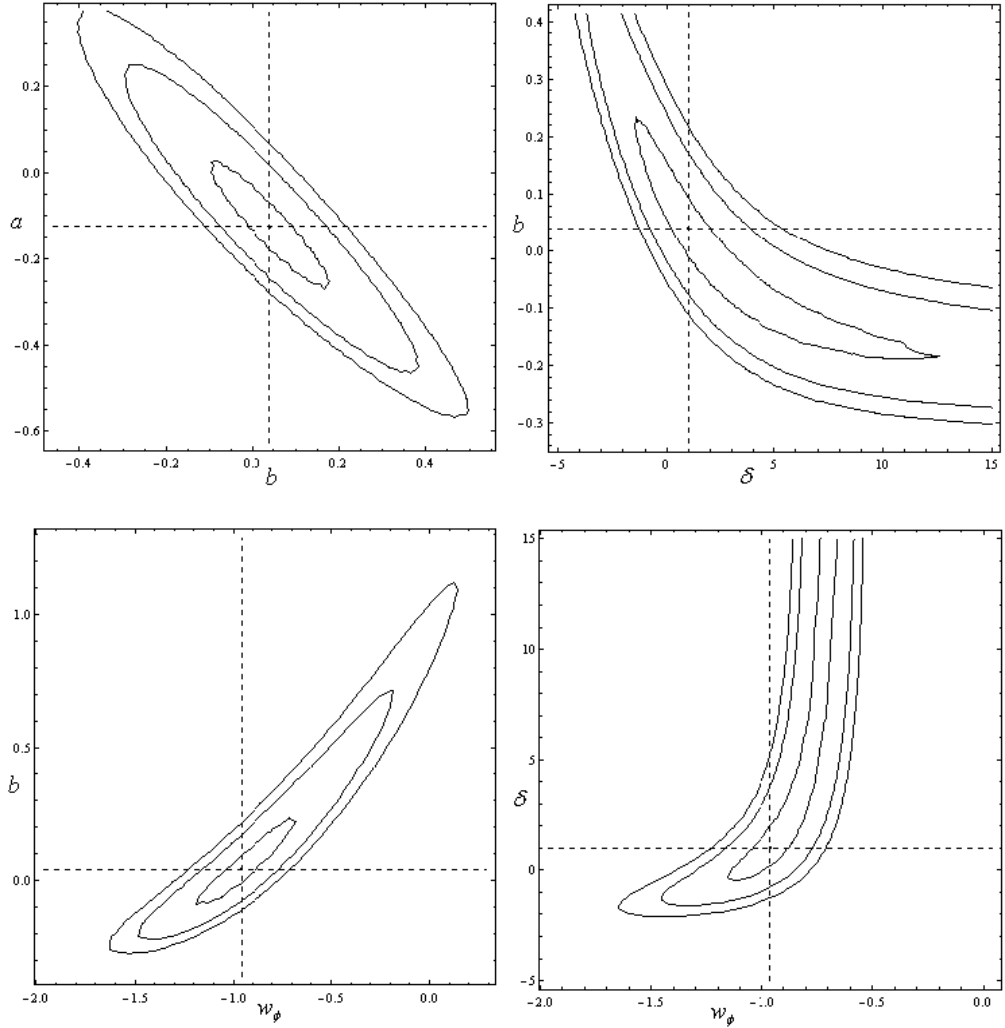

Figure 1. The $\chi^2$-distribution of estimated parameters near the bets fit ones.
Contours display $1\sigma$, $2\sigma$ and $3\sigma$ confidence level.
The dashed lines mean the best fit value of parameters

In ΛCDM model, we can set $w_\varphi = -1$ and recalculate the best-fit values of a, b, δ (Table. 2). It gives

$$\dot{G}/G|_{z=0} = 1.43 \times 10^{-12} \, yr^{-1} , \qquad (47)$$

where we have taken $H_0 = 70 \, km/s \cdot Mpc = 7.15 \times 10^{-11} \, yr^{-1}$. This value is similar to, but a bit greater still than the value in [18].

| parameters | Union($N_{SN} = 307$) + BAO |
|---|---|
| $a$ | $-0.08^{+0.16}_{-0.15}$ |
| $b$ | $-0.02^{+0.18}_{-0.25}$ |
| $\delta$ | $+0.21^{+7.8}_{-2.04}$ |

Table 2. Best fit values of parameters with $w_\varphi = -1$

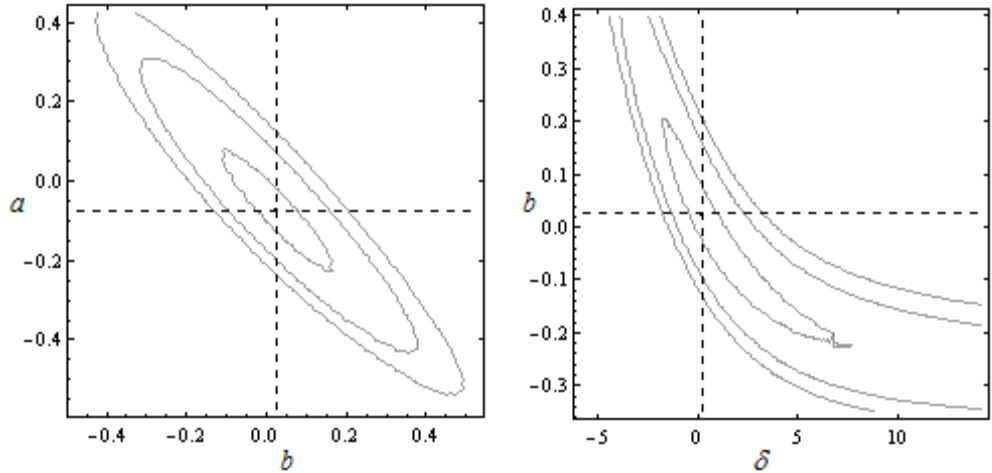

Figure 2. The $\chi^2$-distribution of estimated parameters near the best-fit ones with $w_\phi = -1$.
Contours display $1\sigma$, $2\sigma$ and $3\sigma$ confidence level.
The dashed lines mean the best fit value of parameters

## IV. DISCUSSION AND CONCLUSION

In contrast with [3, 4, 18], in this paper we constructed the generalized model of the scalar-tensor gravity that has the direct coupling between scalar and dark matter and inspected the coupling, using the cosmological dataset of SN and BAO. The model we suggested can include the general relativity and it can examine the scalar-matter coupling if any.

In this paper we applied a new parameterization for the function $F(z) = \dfrac{G_0}{G(z)}$ including a linear term in contrast with [1]. This makes it possible to estimate the time-varying rate of gravitational constant according to (33): $\dot{G}/G|_{z=0}=2.86\times10^{-12}\,yr^{-1}$. This constraint is weaker than the estimation in [23].

On the other hand, as regards the scalar-matter coupling, Table.1 shows a positive coupling parameter $\delta=+1.01$. Guo et al. has given rather negative value $\delta=-0.03$ [4]. The contrary results between ours and [4] seem to stem from the difference of the models and the utilized observation datasets.

Also, we recalculated and gained the time-varying rate of gravitational constant with $w_\varphi=-1$: $\dot{G}/G|_{z=0}=1.43\times10^{-12}\,yr^{-1}$, which is more closer to the range $|\dot{G}/G|\leq 9\times10^{-13}\,yr^{-1}$ in [18], but it is a bit greater.

In a future publication we wish to include the CMB anisotropy data and the combined analysis of SNIa, CMB and BAO datasets is believed to give more conclusive results.